\documentclass[10pt]{article}
\usepackage{graphicx}
\usepackage{dcolumn}
\usepackage{bm}
\usepackage{color}
\usepackage{amssymb}
\usepackage{amsmath}
\usepackage{multicol,times}
\usepackage[all]{xy}
\usepackage{arydshln}

\begin{document}

\title{\textbf{Triadic motifs in the dependence networks of virtual societies}}
\author{
\noindent
Wen-Jie Xie{$^{1,2}$},
Ming-Xia Li{$^{1,2}$},
Zhi-Qiang Jiang{$^{1,3}$}
\& Wei-Xing Zhou{$^{1,2,3}$}
}

\maketitle

\vspace{-5mm}

\noindent
$^1$Departmenent of Mathematics, East China University of Science and Technology, Shanghai 200237, China,
$^2$School of Business, East China University of Science and Technology, Shanghai 200237, China,
$^3$Research Center for Econophysics, East China University of Science and Technology, Shanghai 200237, China

\medskip
\noindent
Correspondence and requests for materials should be addressed to W.X.Z. (wxzhou@ecust.edu.cn).

\vspace{1mm}
\begin{center}
({\em{Received \today; Accepted XXX; Published XXX}})
\end{center}
\vspace{1mm}

\begin{quote}
  {\noindent\textbf{In friendship networks, individuals have different numbers of friends, and the closeness or intimacy between an individual and her friends is heterogeneous. Using a statistical filtering method to identify relationships about who depends on whom, we construct dependence networks (which are directed) from weighted friendship networks of avatars in more than two hundred virtual societies of a massively multiplayer online role-playing game (MMORPG). We investigate the evolution of triadic motifs in dependence networks. Several metrics show that the virtual societies evolved through a transient stage in the first two to three weeks and reached a relatively stable stage. We find that the unidirectional loop motif (${\rm{M}}_9$) is underrepresented and does not appear, open motifs are also underrepresented, while other close motifs are overrepresented. We also find that, for most motifs, the overall level difference of the three avatars in the same motif is significantly lower than average, whereas the sum of ranks is only slightly larger than average. Our findings show that avatars' social status plays an important role in the formation of triadic motifs.
}}
\end{quote}

The structure and social function of friendship networks formed of individuals with friendship ties, are of crucial importance in relation to our understanding of the evolution and behaviors of complex social systems. Conventionally, the construction of friendship networks are implemented by interviews or questionnaires \cite{Currarini-Jackson-Pin-2009-Em,Currarini-Jackson-Pin-2010-PNAS,Ball-Newman-2013-NS}. Usually, the sizes of these networks are not large and the samples are biased. Modern technology provides us alternative ways for the collections of data on social relationships. For instance, the wide use of mobile phones leaves footprints of human's social activities, making it possible to record and analyze very large social networks \cite{Palla-Barabasi-Vicsek-2007-Nature,Onnela-Saramaki-Hyvonen-Szabo-Lazer-Kaski-Kertesz-Barabasi-2007-PNAS,Kumpula-Onnela-Saramaki-Kaski-Kertesz-2007-PRL,Eagle-Penland-Lazer-2009-PNAS,Jo-Pan-Kaski-2011-PLoS1,Jiang-Xie-Li-Podobnik-Zhou-Stanley-2013-PNAS,Kovanen-Kaski-Kertesz-Saramaki-2013-PNAS}. Another important source comes from the virtual worlds resided in computer servers running massively multiplayer online role-playing games (MMORPGs). The availability of big data recorded from MMORPGs potentially enables us to test social and economic hypotheses and theories in large-scale virtual populations \cite{Bainbridge-2007-Science}. In recent years, such kind of data has been investigated from different angles of view \cite{Jiang-Zhou-Tan-2009-EPL,Jiang-Ren-Gu-Tan-Zhou-2010-PA,Thurner-Szell-Sinatra-2012-PLoS1,Szell-Sinatra-Petri-Thurner-Latora-2012-SR,Szell-Thurner-2012-ACS}. In particular, studies on the structure and dynamic evolution of social networks in virtual worlds have unveiled intriguing results \cite{Szell-Lambiotte-Thurner-2010-PNAS,Szell-Thurner-2010-SN,Klimek-Thurner-2013-NJP,Szell-Thurner-2013-SR}.

In virtual worlds, avatars (virtual characters in MMORPGs) interact with each other through their social and economic activities. An avatar can propose to another avatar to make friends when they encounter. If the later accepts the proposal, they become friends and their names appear in their friend lists. Hence, the friendship is reciprocal. We will investigate the friend lists of all avatars in 248 virtual societies of a popular MMORPG (for more information about the data sets, see {\textit{Methods}}). In this MMORPG, the closeness of two avatars that are friends is measured and recorded, which is termed as intimacy. If two friends finish social or economic tasks together, their intimacy will increase. Therefore, the friendship networks are weighted. Just like almost all other social networks, these friendship networks are very heterogeneous in the sense that they have broad degree distributions and that each avatar has diverse intimacies. Indeed, we find that a large proportion of friendship ties have zero intimacy, while many other ties have very large intimacy. It suggests that some friendship ties are more important and significant than others. It is thus necessary to remove insignificant ties from the friendship networks.

The simplest and natural way is to set a threshold for the intimacy \cite{Xie-Li-Jiang-Tan-Podobnik-Zhou-Stanley-2014-PNAS}. If the intimacy of two friends is less than the threshold, the two avatars are not treated as friends. In this work, we adopt an alternative and more systematic way of filtering proposed by Serrano et al. \cite{Serrano-Boguna-Vespignani-2009-PNAS}. The idea is to statistically validate the links of each avatar by identifying which of her links carry disproportionate fraction of the weights. Links that deviate significantly a preset null model are kept to form the ``multiscale backbone'' of the original network. A significant link between an avatar $i$ and her friend $j$ means that the link plays an important role among all her links and we can argue that avatar $i$ depends on her friend $j$. This results in a directed link from $i$ to $j$. We call the resulting network as dependence network, which is directed.

In this work, we investigate the evolution of triadic motifs in the dependence networks among virtual societies. We find that some motifs occur more than random while other motifs appear less than random. We also find that the avatars forming these triadic motifs often have similar levels. These features are related to the evolution of the virtual societies.

\section*{Results}

\begin{figure}[hb]
  \centering
  \includegraphics[width=0.95\textwidth]{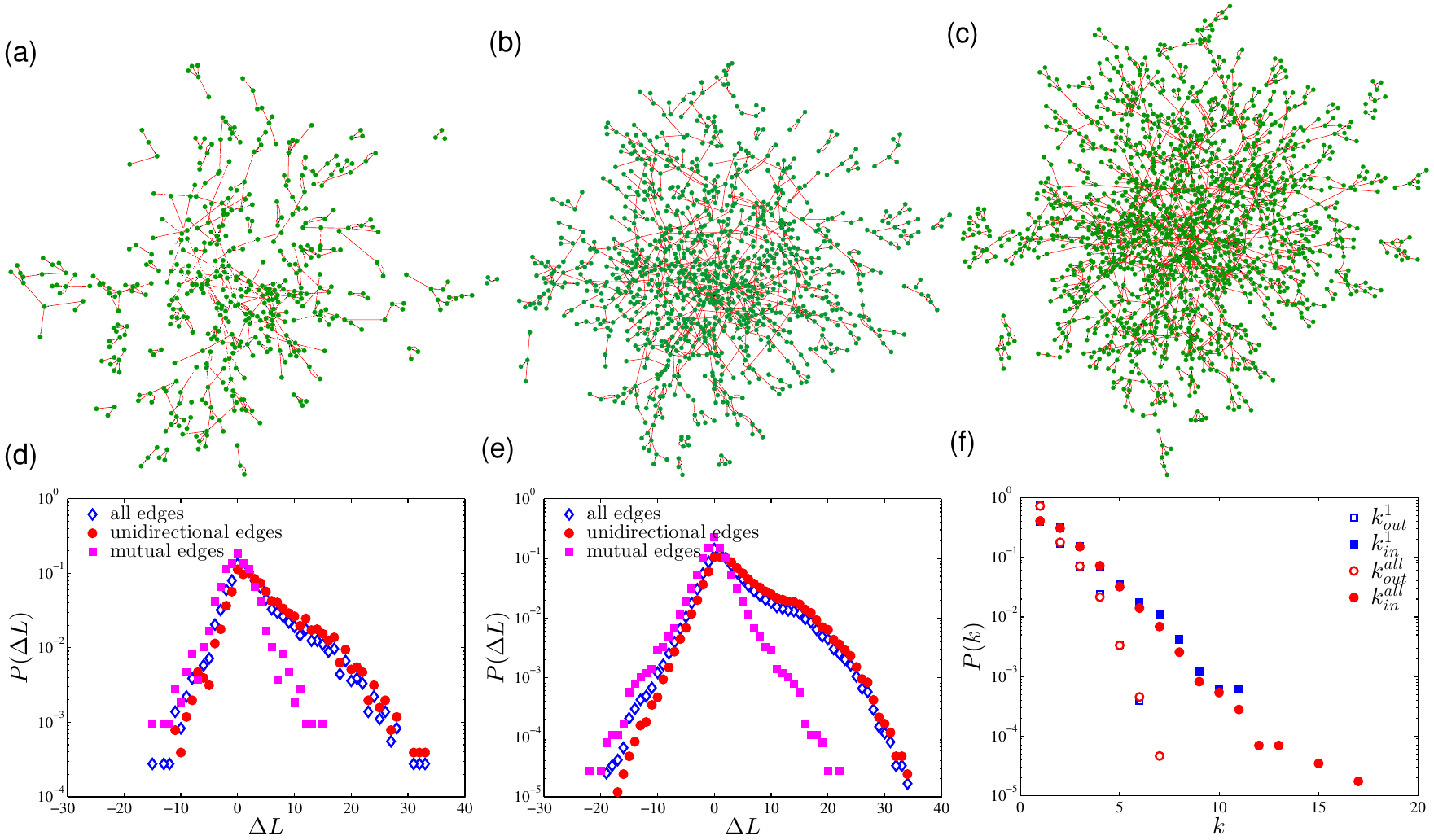}
  \caption{{\textbf{Illustration of dependence networks, degree distributions and distribution of level differences.}} (a) The dependence network of a virtual society on day 2. (b) The dependence network of the same virtual society on day 5. (c) The dependence network of the same virtual society on day 30. (d) Distributions of $\Delta{L_{ij}}$ for all links, for nonreciprocal links (link $i{\to}j$ exists while link $j{\to}i$ does not exist), and for reciprocal links (both links $i{\to}j$ and $j{\to}i$ exist) on day 30 for the same virtual society. (e) Distributions of $\Delta{L_{ij}}$ for all links, for nonreciprocal links, and for reciprocal links on day 30 for all virtual society. (f) In-degree distribution and out-degree distribution for the same virtual society and for all societies.}
  \label{Fig:1}
\end{figure}

\noindent{\textbf{Illustration of dependence networks, degree distributions and distribution of level differences.}}
For each network (a society on a given day), we construct its dependence network (see {\textit{Methods}}). In Fig.~\ref{Fig:1}(a) to (c), we illustrate the evolution of the dependence network in a virtual society on three days. On the second day, the number of edges is low and hence the number of triadic motifs is also low. It is consistent with the fact that significant and stable friendship relations are still infant and under development. Such dependence relationship increases with the evolution of the virtual society because collaborations among avatars increase as well as their intimacies. In addition, a link present in an early dependence network may disappear in the later network.

For a link $i\to{j}$, we define the difference of levels of the two avatars as follows
\begin{equation}
 \Delta{L}_{ij} = L_j -L_i,
 \label{Eq:dL}
\end{equation}
where $L_i$ and $L_j$ are the levels of avatars $i$ and $j$. Figure \ref{Fig:1}(d) plots the distribution of $\Delta{L}$ for all links, as well as nonreciprocal links and reciprocal links on day 30 for the same virtual society as in Fig.~\ref{Fig:1}(a-c), while Fig.~\ref{Fig:1}(e) shows the distributions for all societies. We find that the left part of each distribution can be approximated by an exponential distribution. The distribution for reciprocal links is symmetric in reference to $\Delta{L}=0$ as expected because both $\Delta{L}_{ij}$ and $\Delta{L}_{ji}$ are counted. The average proportion of reciprocal links increases from 10\% to 23\%. However, for nonreciprocal links, the right tail of the distribution is much fatter than the left tail. On average, about 85\% of the nonreciprocal links run from low-level avatars to high-level avatars and this ratio increases along time from about 82\% to 86\%.

Figure \ref{Fig:1}(f) shows the distributions of in-degrees and out-degrees of the dependence network in Fig.~\ref{Fig:1}(c). The in-degree distribution decays exponentially, while the out-degree distribution decays faster than an exponential. It is intriguing to observe that the in-degree distribution is much fatter than the out-degree distribution. This observation is rational and can be explained in a mechanical way. Usually, a powerful avatar is able to help many less powerful avatars and play a relative important role in the friendship lists of those less powerful avatars. However, it is difficult for an avatar to have relatively similar intimacies from many different friends.

\begin{figure}[tb]
  \centering
  \includegraphics[width=18cm]{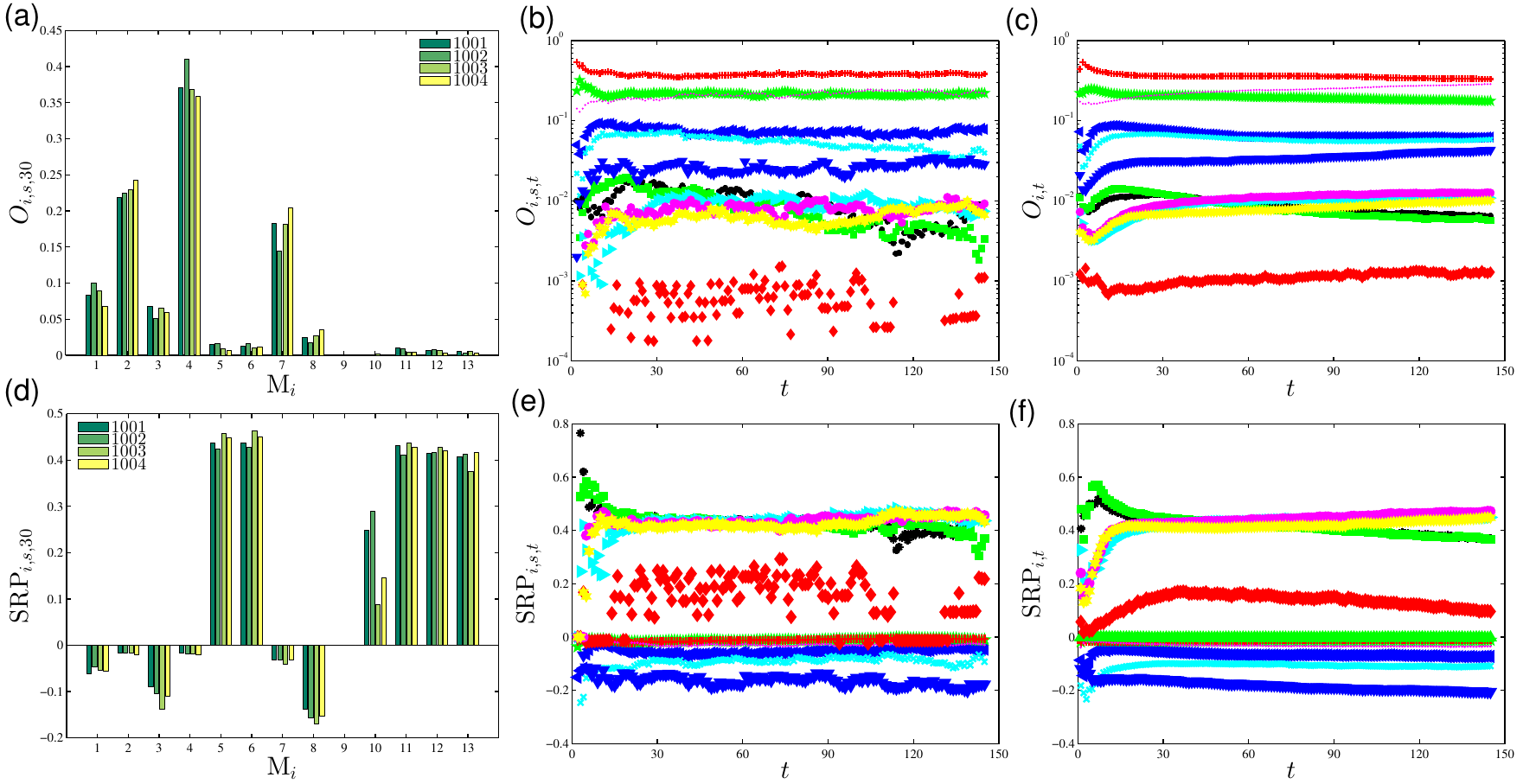}
  \caption{{\textbf{Distribution of triadic motifs.}}
  (a) Occurrence frequencies $O_{i,s,30}$ of the 13 motifs on the 30th day ($t=30$) for four virtual societies.
  (b) Evolution of occurrence frequencies $O_{i,s,t}$ of the 13 triadic motifs for a typical virtual society $s$. The occurrence frequency of motif 9 is zero along time, that is $O_{9,s,t}=0$.
  (c) Evolution of average occurrence frequencies $O_{i,t}=\frac{1}{S}\sum_{s=1}^S O_{i,s,t}$ of the 13 triadic motifs over all virtual society. The average occurrence frequency of motif 9 is zero along time, that is $O_{9,t}=0$.
  (d) Significance  ${\rm{SRP}}_{i,s,30}$ of the 13 motifs on the 30th day ($t=30$) for four virtual societies.
  (e) Evolution of significance  ${\rm{SRP}}_{i,s,t}$ of the 13 triadic motifs for a typical virtual society $s$.
  (f) Evolution of average significance ${\rm{SRP}}_{i,t}=\frac{1}{S}\sum_{s=1}^S {\rm{SRP}}_{i,s,t}$ of the 13 triadic motifs over all virtual society.
  The correspondence between the colourful symbols in plots (b), (c), (e) and (f) and motifs are given in Fig.~\ref{Fig:5}.}
  \label{Fig:2}
\end{figure}

\noindent{\textbf{Distribution and evolution of motif occurrence frequency.}}
Let $N_{i,s,t}$ denote the number of motif $i$ in society $s$ on day $t$. The relative occurrence frequency of motif $i$ in virtual society $s$ on day $t$ is
\begin{equation}
  O_{i,s,t}=\frac{N_{i,s,t}}{\sum_{j=1}^{13} N_{j,s,t}}.
\end{equation}
Figure \ref{Fig:2}(a) shows the distribution of relative occurrence frequencies of the 13 motifs on day 30, $O_{i,s,30}$, for four typical virtual societies. We find that the distributions are quite similar for different societies and open motifs (${\rm{M}}_1$, ${\rm{M}}_2$, ${\rm{M}}_3$, ${\rm{M}}_4$, ${\rm{M}}_7$ and ${\rm{M}}_8$) have higher occurrence frequency than close motifs. We also find that the occurrence frequency of motif 9 is $O_{9,s,30}=0$. It suggests that the situation in which avatar $a$ depends on avatar $b$, avatar $b$ depends on avatar $c$, and avatar $c$ depends on avatar $a$ is unlikely to appear. These results are also observed for other virtual societies on other days, as shown in Fig.~\ref{Fig:2}(b) for a typical society. It is found that the occurrence frequency experienced a transient stage in the first one or two weeks and then became relatively stable. For instance, the curve for ${\rm{M}}_4$ on the top of the plot decreases fast in the first few days and becomes almost horizontal, while many other curves exhibit an increasing trend in the first few days and then decrease slowly or become almost horizontal. We also show in Fig.~\ref{Fig:2}(c) the evolution of the average occurrence frequency over all societies on day $t$:
\begin{equation}
  O_{i,t}=\frac{1}{S}\sum_{s=1}^S O_{i,s,t}.
\end{equation}
We confirm that there is no ${\rm{M}}_9$ in all dependence networks.

\noindent{\textbf{Significance of motif occurrences.}}
Some motifs are more likely to appear than other motifs in networks, even if there are no social factors in the formation of networks. Hence, we determine the occurrence significance of motifs by the $z$-score
\begin{equation}
  Z_{i,s,t}=\frac{N_{i,s,t}-\langle N_{i,s,t,{\rm{rand}}}\rangle}{\sigma(N_{i,s,t,{\rm{rand}}})},
\end{equation}
where $N_{i,{\rm{rand}}}$ is the occurrence number of ${\rm{M}}_i$ in a randomized network from the dependence network of society $s$ on day $t$, and $\langle N_{i,{\rm{rand}},s,t}\rangle$ and $\sigma(N_{i,{\rm{rand}},s,t})$ are the mean and standard deviation of $N_{i,{\rm{rand}},s,t}$ over 100 randomized networks \cite{Milo-Itzkovitz-Kashtan-Levitt-ShenOrr-Ayzenshtat-Sheffer-Alon-2004-Science,Milo-Kashtan-Itzkovitz-Newman-Alon-2004-XXX}.

The results of the dependence network in a typical virtual society on day 30 are shown in Table \ref{Tb:Zi}. Again, we find that the 13 motifs can be classified into three groups. The first group contains ${\rm{M}}_9$, where $N_{i=9,s,t}=0$. Although ${\rm{M}}_9$ was not detected in real networks, it appears occasionally in random networks. The second group contains all open motifs ${\rm{M}}_1$, ${\rm{M}}_2$, ${\rm{M}}_3$, ${\rm{M}}_4$, ${\rm{M}}_7$ and ${\rm{M}}_8$, where $N_{i,s,t}<\langle N_{i,s,t,{\rm{rand}}}\rangle$. The third group contains other close motifs ${\rm{M}}_5$, ${\rm{M}}_6$, ${\rm{M}}_{10}$, ${\rm{M}}_{11}$, ${\rm{M}}_{12}$ and ${\rm{M}}_{13}$, where $N_{i,s,t}>\langle N_{i,s,t,{\rm{rand}}}\rangle$. It is worthy stressing that ${\rm{M}}_{13}$ appears in real networks but not in random networks. We find that all the standard deviations are less than 1 and the $z$-scores are significantly different from zero. Therefore, when compared with random networks, open motifs are less likely to occur while close motifs (except ${\rm{M}}_9$) are more likely to appear in real dependent networks. For close motifs except ${\rm{M}}_{13}$, the standard deviation is greater than the mean. In addition, there are five close motifs (${\rm{M}}_5$, ${\rm{M}}_6$, ${\rm{M}}_{10}$, ${\rm{M}}_{11}$, ${\rm{M}}_{12}$) having $\langle N_{i,{\rm{rand}},s,t}\rangle=0.10$ and $\sigma(N_{i,{\rm{rand}},s,t})=0.30$. It can be understood due to the fact that, for each motif $i$ ($i=5,6,10,11,12$), there are 10 random networks (out of 100) having $N_{i,{\rm{rand}},s,t}=1$.

\begin{table}[t]
  \caption{{\textbf{Statistical significance of motif occurrences for a typical virtual society on day 30.}} Without loss of clarity, we have dropped the subscripts $s$ and $t$ in the variables. The motifs can be classified into three groups: (1) ${\rm{M}}_9$ with zero occuurence; (2) Open motifs ${\rm{M}}_1$, ${\rm{M}}_2$, ${\rm{M}}_3$, ${\rm{M}}_4$, ${\rm{M}}_7$ and ${\rm{M}}_8$ that appear significantly less than random. (3) Close motifs ${\rm{M}}_5$, ${\rm{M}}_6$, ${\rm{M}}_{10}$, ${\rm{M}}_{11}$, ${\rm{M}}_{12}$ and ${\rm{M}}_{13}$ that appear significantly more than random. }
  \smallskip
  \label{Tb:Zi}
  \centering
  \begin{tabular}{ccrcrcrc}
  \hline\hline
  $i$   & ${\rm{M}}_i$ & $O_{i}$ & $N_{i,\rm{real}}$  & $\langle{N_{i,\rm{rand}}}\rangle$ & $\sigma(N_{i,\rm{rand}})$ & $Z_i$ & $p$-value\\\hline
  9  &  \begin{xy}
          \POS (0,3) *\cir<2pt>{} ="a", (-4,-2)*\cir<2pt>{} ="b", (4,-2)*\cir<2pt>{} ="c"
          \POS "a" \ar @{->} "b"
          \POS "a" \ar @{<-} "c"
          \POS "b" \ar @{->} "c"
        \end{xy}
     & 0.00000  &  0  &  0.10  &  0.30  &  -0.27&  1.00\\\hdashline
  1  &  \begin{xy}
          \POS (0,3) *\cir<2pt>{} ="a", (-4,-2)*\cir<2pt>{} ="b", (4,-2)*\cir<2pt>{} ="c"
                    \POS "a" \ar @{->} "b"
                    \POS "a" \ar @{->} "c"
                \end{xy}
     & 0.08358  &  477  &  622.60  &  0.50  &  -273.78&  1.00\\
  2  &  \begin{xy}
          \POS (0,3) *\cir<2pt>{} ="a", (-4,-2)*\cir<2pt>{} ="b", (4,-2)*\cir<2pt>{} ="c"
          \POS "a" \ar @{<-} "b"
          \POS "a" \ar @{->} "c"
        \end{xy}
     & 0.21850  &  1247  &  1339.41  &  0.00  &  -95.85&  1.00\\
  3  &  \begin{xy}
          \POS (0,3) *\cir<2pt>{} ="a", (-4,-2)*\cir<2pt>{} ="b", (4,-2)*\cir<2pt>{} ="c"
          \POS "a" \ar @{<->} "b"
          \POS "a" \ar @{->} "c"
        \end{xy}
     & 0.06746  &  385  &  571.70  &  0.80  &  -247.95&  1.00\\
  4  &  \begin{xy}
          \POS (0,3) *\cir<2pt>{} ="a", (-4,-2)*\cir<2pt>{} ="b", (4,-2)*\cir<2pt>{} ="c"
          \POS "a" \ar @{->} "c"
          \POS "b" \ar @{->} "c"
        \end{xy}
     & 0.37112  &  2118  &  2278.60  &  0.60  &  -265.61&  1.00\\
  7  &  \begin{xy}
          \POS (0,3) *\cir<2pt>{} ="a", (-4,-2)*\cir<2pt>{} ="b", (4,-2)*\cir<2pt>{} ="c"
          \POS "a" \ar @{<->} "b"
          \POS "a" \ar @{<-} "c"
        \end{xy}
     & 0.18311  &  1045  &  1201.70  &  0.60  &  -254.97&  1.00\\
  8  &  \begin{xy}
          \POS (0,3) *\cir<2pt>{} ="a", (-4,-2)*\cir<2pt>{} ="b", (4,-2)*\cir<2pt>{} ="c"
          \POS "a" \ar @{<->} "b"
          \POS "a" \ar @{<->} "c"
        \end{xy}
     & 0.02541  &  145  &  270.90  &  0.30  &  -461.82&  1.00\\\hdashline
  5  &  \begin{xy}
          \POS (0,3) *\cir<2pt>{} ="a", (-4,-2)*\cir<2pt>{} ="b", (4,-2)*\cir<2pt>{} ="c"
          \POS "a" \ar @{->} "b"
          \POS "a" \ar @{->} "c"
          \POS "b" \ar @{->} "c"
        \end{xy}
     & 0.01542  &  88  &  0.30  &  0.50  &  171.89&  0.00\\
  6  &  \begin{xy}
          \POS (0,3) *\cir<2pt>{} ="a", (-4,-2)*\cir<2pt>{} ="b", (4,-2)*\cir<2pt>{} ="c"
          \POS "a" \ar @{<->} "b"
          \POS "a" \ar @{->} "c"
          \POS "b" \ar @{->} "c"
        \end{xy}
     & 0.01279  &  73  &  0.10  &  0.30  &  227.26&  0.00\\
  10 & \begin{xy}
          \POS (0,3) *\cir<2pt>{} ="a", (-4,-2)*\cir<2pt>{} ="b", (4,-2)*\cir<2pt>{} ="c"
          \POS "a" \ar @{->} "b"
          \POS "a" \ar @{<->} "c"
          \POS "b" \ar @{->} "c"
        \end{xy}
     & 0.00088  &  5  &  0.10  &  0.30  &  19.23&  0.00\\
  11 & \begin{xy}
          \POS (0,3) *\cir<2pt>{} ="a", (-4,-2)*\cir<2pt>{} ="b", (4,-2)*\cir<2pt>{} ="c"
          \POS "a" \ar @{<-} "b"
          \POS "a" \ar @{<->} "c"
          \POS "b" \ar @{->} "c"
        \end{xy}
     & 0.01016  &  58  &  0.10  &  0.30  &  212.43&  0.00\\
  12 & \begin{xy}
          \POS (0,3) *\cir<2pt>{} ="a", (-4,-2)*\cir<2pt>{} ="b", (4,-2)*\cir<2pt>{} ="c"
          \POS "a" \ar @{<->} "b"
          \POS "a" \ar @{<->} "c"
          \POS "b" \ar @{->} "c"
        \end{xy}
    & 0.00631  &  36  &  0.10  &  0.30  &  131.74&  0.00\\
  13 & \begin{xy}
          \POS (0,3) *\cir<2pt>{} ="a", (-4,-2)*\cir<2pt>{} ="b", (4,-2)*\cir<2pt>{} ="c"
          \POS "a" \ar @{<->} "b"
          \POS "a" \ar @{<->} "c"
          \POS "b" \ar @{<->} "c"
        \end{xy}
    &  0.00526  &  30  &  0.00  &  0.00  &  0.00&  0.00\\
  \hline\hline
  \end{tabular}
\end{table}

Because the standard deviation can be very small or even zero when a subgraph is unlikely to occur in a network, the estimated $z$-score fluctuates a lot and may diverge. An alternative measure can be adopted to assess the occurrence significance of motifs \cite{Milo-Itzkovitz-Kashtan-Levitt-ShenOrr-Ayzenshtat-Sheffer-Alon-2004-Science}. We calculate the abundance of ${\rm{M}}_i$ relative to random networks as follows
\begin{equation}
  \Delta_{i,s,t}=\frac{N_{i,s,t}-\langle N_{i,s,t,{\rm{rand}}}\rangle}{N_{i,s,t}+\langle N_{i,s,t,{\rm{rand}}}\rangle+\varepsilon},
\end{equation}
where $\varepsilon=4$ is to ensure that $|\Delta_i|$ is not too large when ${\rm{M}}_i$ appears very few times in both real and random networks, i.e., both $N_{i,s,t}$ and $\langle N_{i,s,t,{\rm{rand}}}\rangle$ are close to zero \cite{Milo-Itzkovitz-Kashtan-Levitt-ShenOrr-Ayzenshtat-Sheffer-Alon-2004-Science}. This special case happens for ${\rm{M}}_9$. We then compute the normalized abundance as follows
\begin{equation}
  {\rm{SRP}}_{i,s,t}=\frac{\Delta_{i,s,t}}{\left(\sum \Delta_{i,s,t}^2\right)^{1/2}},
\end{equation}
which is called the subgraph ratio profile (SRP) \cite{Milo-Itzkovitz-Kashtan-Levitt-ShenOrr-Ayzenshtat-Sheffer-Alon-2004-Science}. We can also calculate the average SRP over all virtual societies:
\begin{equation}
  {\rm{SRP}}_{i,t}=\frac{1}{S}\sum_{s=1}^S {\rm{SRP}}_{i,s,t}.
\end{equation}
The average procedure is meaningful because different societies exhibit similar motif profiles.

In Fig.~\ref{Fig:2}(d), we show the subgraph ratio profile ${\rm{SRP}}_{i,s,t}$ for the four societies in Fig.~\ref{Fig:2}(a) on day 30. The three groups identified in Table \ref{Tb:Zi} are also observed. For the first group, we have ${\rm{SRP}}_{i=9,s,t}=0$. For the second group, we have ${\rm{SRP}}_{i,s,t}<0$. For the third group, we have ${\rm{SRP}}_{i,s,t}>0$. We further find that the absolute SRPs for the open motifs in the second group are less than that for the close motifs in the third group. In Fig.~\ref{Fig:2}(e) and (f), we also show the evolution of subgraph ratio profiles ${\rm{SRP}}_{i,s,t}$ for the same society as in Fig.~\ref{Fig:2}(b) and the average subgraph ratio profiles ${\rm{SRP}}_{i,t}$ over all societies. Again, we observe a transient stage and a relative stable stage.

\begin{figure}[htb]
  \centering
  \includegraphics[width=18cm]{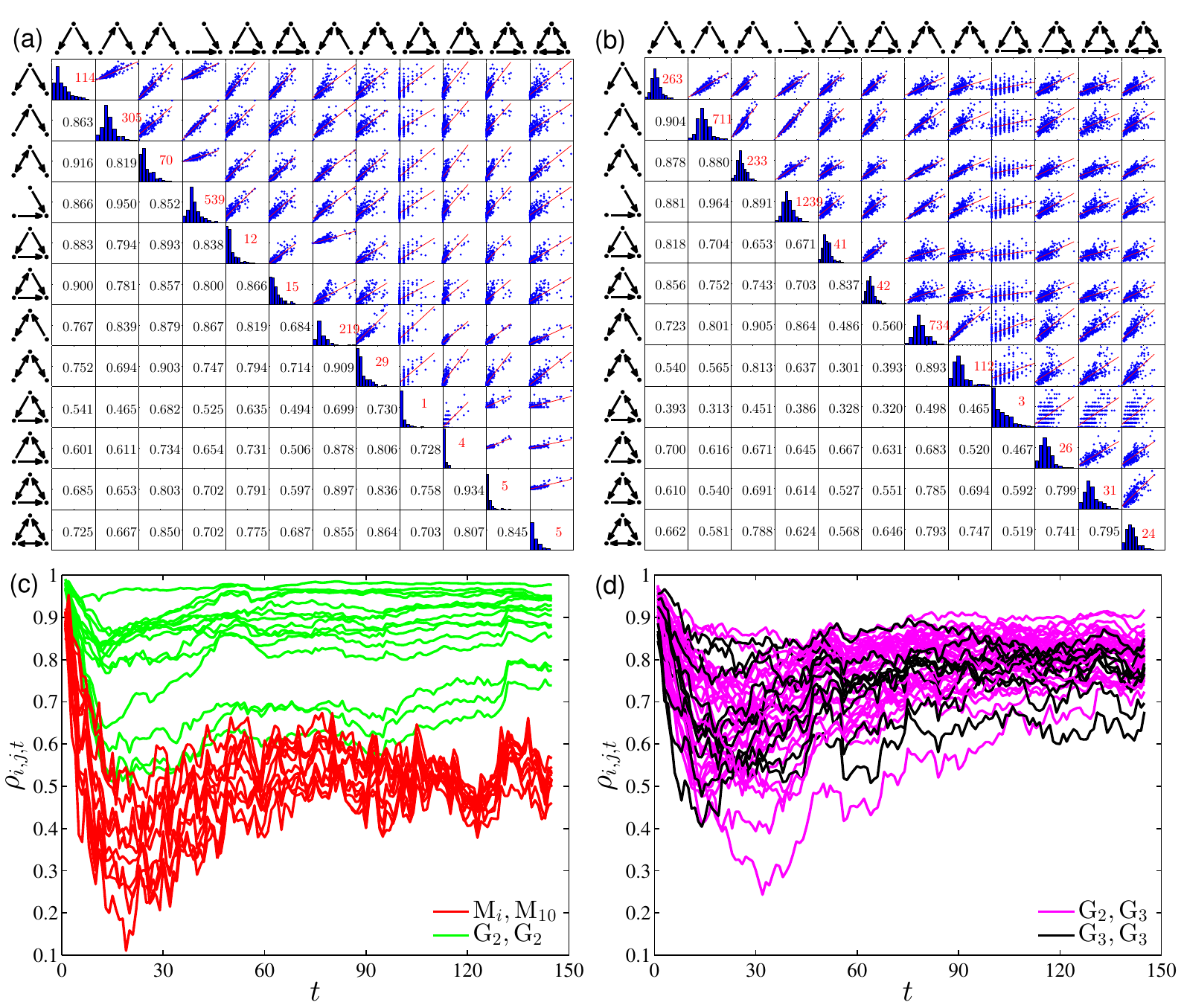}
  \caption{{\textbf{Correlation of motif counts in 248 dependence networks.}} (a) The panels in the upper triangle are scatter plots of motif counts between 13 motif counts on day 7. There are no ${\rm{M}}_9$ motifs detected in the dependence networks. The correlation coefficients of motifs counts are presented in the lower triangular panels. Each diagonal panel shows the count distribution of these motifs and also the average motif count. (b) Same as (a) for day 30. (c) Evolution of motif count correlation coefficient $\rho_{i,j,t}$ between ${\rm{M}}_10$ and other motifs (red curves) and between open motifs in the second group (green curves). (d) Evolution of motif count correlation coefficient $\rho_{i,j,t}$ between close motifs in the third group (black curves) and between open motifs in the second group and close motifs in the third group (magenta curves).}
  \label{Fig:3}
\end{figure}

\noindent{\textbf{Correlation between motif counts.}}
We perform motif count analysis to study the correlation between motif counts in dependence networks, similar as in human cell-specific transcription factor regulatory networks \cite{Tran-Choi-Zhang-2013-NatComm}. Specifically, for a given day $t$, we plot $N_{i,s,t}$ against $N_{j,s,t}$ for different society $s$, where each $s$ gives a point. Panels (a) and (b) of Fig.~\ref{Fig:3} show the correlation plots for $t=7$ and $t=30$, respectively, where ${\rm{M}}_9$ is not included because $N_{i=9,s,t}=0$. For each motif, the distribution of the motif count in the diagonal is unimodal with the peak close to the mean value. We find that the motif counts exhibit linear correlations. On average, the correlation coefficients $\rho_{ij}$ between ${\rm{M}}_{10}$ and other motifs are relatively small, due to the fact that the average count of ${\rm{M}}_{10}$ is small with relatively large fluctuations.

In Fig.~\ref{Fig:3}(c) and (d), we plot the evolution of correlation coefficient $\rho_{i,j,t}$ for all the pairs of motifs. The correlation coefficient $\rho_{i,j,t}$ decreases in the first two or three weeks and then increases to a relatively stable state. The count correlations between open motifs are relatively the strongest for most pairs, while the correlations between ${\rm{M}}_{10}$ and other motifs are relatively the weakest.

The characteristics of these correlations are closely related to the results in Fig.~\ref{Fig:2}(a)-(c). It suggests that the local structure of dependence networks in diverse virtual societies exhibits universal evolution patterns. The relative ratio of motif counts $N_{i,s,t}/N_{j,s,t}$ between ${\rm{M}}_i$ and ${\rm{M}}_j$ converges to certain constant along the evolution of virtual societies.

\noindent{\textbf{Avatar levels in motifs.}}
Abnormal occurrence of motifs is often in relation to certain features of the involving individuals in complex social networks \cite{Milo-ShenOrr-Itzkovitz-Kashtan-Chklovskii-Alon-2002-Science,Milo-Itzkovitz-Kashtan-Levitt-ShenOrr-Ayzenshtat-Sheffer-Alon-2004-Science,Kovanen-Karsai-Kaski-Kertesz-Saramaki-2011-JSM,Jiang-Xie-Xiong-Zhang-Zhang-Zhou-2013-QFL,Kovanen-Kaski-Kertesz-Saramaki-2013-PNAS}. We now investigate the levels of avatars in triadic motifs. The levels of avatars are the most important trait, because levels reflect the power of attack, defence, and production of avatars.

Denote the levels of the three avatars in ${\rm{M}}_i$ by $L_{i,1}$, $L_{i,2}$ and $L_{i,3}$. The sum of avatar levels is
\begin{equation}
  l_i=L_{i,1}+L_{i,2}+L_{i,3}.
\end{equation}
The mean value of level sums of the $N_{i,s,t}$ motifs in society $s$ on day $t$ can be calculated as follows,
\begin{equation}
  l_{i,s,t}=\frac{1}{N_{i,s,t}}\sum_{k=1}^{N_{i,s,t}}l_i^{(k)};
\end{equation}
Similarly, we define the difference of avatar levels as
\begin{equation}
  \Delta l_i=|L_{i,1}-L_{i,2}|+|L_{i,2}-L_{i,3}|+|L_{i,3}-L_{i,1}|=2\left[\max\{L_{i,1},L_{i,2},L_{i,3}\}-\min\{L_{i,1},L_{i,2},L_{i,3}\}\right],
\end{equation}
and calculate its mean as follows
\begin{equation}
  \Delta l_{i,s,t}=\frac{1}{N_{i,s,t}}\sum_{k=1}^{N_{i,s,t}}\Delta l_i^{(k)}.
\end{equation}

\begin{figure}[htb]
  \centering
  \includegraphics[width=12cm]{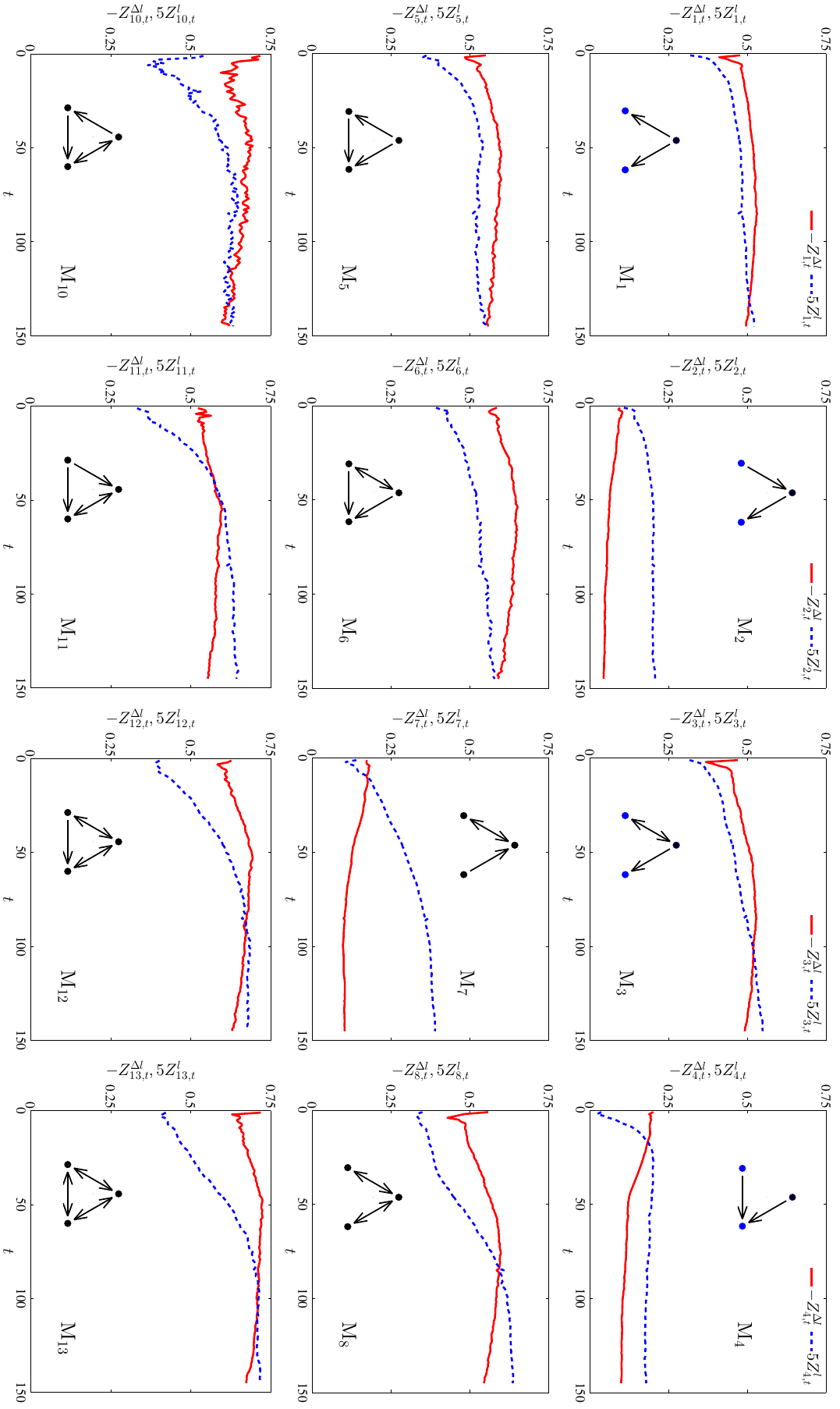}
  \caption{{\textbf{Evolution of $z$-scores of the average level sum and level difference of avatars in motifs.}} Each plot corresponds to a motif. For clarity, the $Z_{i,t}^{l}$ values has been multiplied by a factor of 5.}
  \label{Fig:4}
\end{figure}

To estimate if the sum and difference of avatar levels are significantly different from random networks (see {\textit{Methods}}), we calculate the means of these two metrics for random networks, which are denoted by $\langle l_{i,s,t,{\rm{rand}}}\rangle$ for level sums and $\langle \Delta l_{i,s,t,{\rm{rand}}}\rangle$ for level difference. The $z$-scores of the two metrics can be obtained as follows
\begin{equation}
 Z_{i,s,t}^{l}=\frac{l_{i,s,t}-\langle l_{i,s,t,{\rm{rand}}}\rangle}{l_{i,s,t}+\langle l_{i,s,t,{\rm{rand}}}\rangle}
\end{equation}
and
\begin{equation}
 Z_{i,s,t}^{\Delta l}=\frac{\Delta l_{i,s,t}-\langle \Delta l_{i,s,t,{\rm{rand}}}\rangle}{\Delta l_{i,s,t}+\langle \Delta l_{i,s,t,{\rm{rand}}}\rangle}.
\end{equation}
The averaged $z$-scores over all virtual societies are computed as follows
\begin{equation}
  Z_{i,t}^{l}=\frac{1}{S}\sum_{s=1}^S  Z_{i,s,t}^{l}
\end{equation}
and
\begin{equation}
  Z_{i,t}^{\Delta l}=\frac{1}{S}\sum_{s=1}^S  Z_{i,s,t}^{\Delta l}.
\end{equation}
Figure \ref{Fig:4} shows the evolution of average $z$-scores $Z_{i,t}^{l}$ and $Z_{i,t}^{\Delta l}$ for all motifs excluding ${\rm{M}}_9$.

It is found that the $Z_{i,t}^{l}$ values are positive. It indicates that the average level sum among avatars in the motifs is larger than the average level of all avatars. This observation can be explained by the fact that avatars in motifs often have relatively large intimacies and thus are more active in virtual societies. More activities may result in a relatively faster increase of levels. The $Z_{i,t}^{l}$ curves have an overall increasing trend. Most curves increase relatively rapid in the first few weeks and then slow down, while some curves (say, ${\rm{M}}_1$ and ${\rm{M}}_3$) exhibit a continuous increase. It again indicates that, with the evolution of virtual societies, avatars in the motifs are even more active than other avatars. However, we notice that all the $Z_{i,t}^{l}$ values are less than 0.15, suggesting that the average level of avatars is not a determining factor in the formation of dependence motifs.

On the contrary, all the $Z_{i,t}^{\Delta l}$ values are negative, indicating that the average level difference among avatars in the motifs is smaller than that of all avatars. We can identify two groups of motifs with distinct behaviors. Three motifs (${\rm{M}}_2$, ${\rm{M}}_4$ and ${\rm{M}}_7$) have relative small absolute $Z_{i,t}^{\Delta l}$ values and the absolute $z$-scores decrease along time. These motifs are less significant than others, as shown by the SRP profiles in Fig.~\ref{Fig:2}(d). Nine other motifs exhibit large absolute $Z_{i,t}^{\Delta l}$ values and the absolute $z$-scores increase first and then decrease. In the early stage of the virtual societies, avatars are more enthusiastic to collaborate in their implementation of tasks. Although avatars are apt to make friends and collaborate with higher-level avatars, such kind of imbalanced teams with high level differences are not common unless that the higher-level avatars are very altruistic. Even such imbalanced teams form, the level difference among avatars will decrease because low-level avatars benefit more in such collaborations. This results in the increase of the absolute $Z_{i,t}^{\Delta l}$ values. Several months after the generation of a virtual society, many avatars become less active or even inactive such that their levels increase relatively slowly, whereas the dependence motifs are persistent. The absolute $Z_{i,t}^{\Delta l}$ values decrease.

\section*{Discussion}

We have studied the friendship networks of hundreds of virtual societies. In friendship networks, the distributions of degree and weight (measuring the closeness or intimacy between two friends) are heterogeneous. For a given avatar $i$, it is often true that some friends are more helpful than others. If the intimacy between $i$ and her friend $j$ is significantly larger than other intimacies between $i$ and her other friends, we argue that $i$ depends on $j$. Based on a filtering method \cite{Serrano-Boguna-Vespignani-2009-PNAS}, we construct dependence networks as the ``multiscale backbone'' of the original friendship networks. The dependence networks allow us to study the reciprocal and altruistic behaviors among avatars by ignoring insignificant ties that bear marginal social information.

We focused on the evolution of triadic motifs in dependence networks. Our findings show that the occurrences of motifs evolved to a stable state and are persistent. A very interesting result is that the level difference of avatars in motifs is an important factor in the formation of dependence networks, whereas the level sum of avatars are not. This finding is the same as the friendship networks of US students \cite{Ball-Newman-2013-NS}. It suggests that virtual societies share certain common dynamics as our real society. Our work also highlights the scientific potential of virtual societies in the study of human behaviors \cite{Bainbridge-2007-Science}.

\section*{Methods}

\noindent{\textbf{Data description.}}
We use a huge database recorded from $K=124$ servers of a popular Massively Multiplayer Online Role-Playing Game (MMORPG) in China. In a virtual world residing in a server there are two opposing societies. Two avatars $i$ and $j$ can make friends, and a measure of closeness $I_{ij}$ called intimacy is assigned to the friendship link. The intimacy $I_{ij}$ can increase according to the collaborative activities of $i$ and $j$ if they belong to a same society; Otherwise, $I_{ij}$ remains zero if $i$ and $j$ belong to two different societies. Hence the friendship networks of the two camps are essentially separate. Each avatar can maintain a friendship list, denoted as ${\cal{F}}_i$ for avatar $i$. The friendship relation is symmetric: if $i\in{\cal{F}}_j$, then $j\in{\cal{F}}_i$.

\noindent{\textbf{Construction of dependence network.}}
For a friendship network, we can construct a dependent network by filtering out insignificant edges based on a method proposed by Serrano {\textit{et al}} \cite{Serrano-Boguna-Vespignani-2009-PNAS}. We can construct a directed weighted network $W$ whose element $w_{ij}$ is the relative intimacy of avatar $j$ in reference to all friends of avatar $i$:
\begin{equation}
   w_{ij} =  \frac{I_{ij}}{\sum_{j\in{\cal{F}}_i}I_{ij}},
    \label{Eq:XCB_InvestPatt}
\end{equation}
where $w_{ij}\neq{w}_{ji}$. Following Ref.~\cite{Serrano-Boguna-Vespignani-2009-PNAS}, a directed link $i\to{j}$ is significant at the level of $\alpha$ if
\begin{equation}
   \alpha_{ij} =  1-(k_i-1)\int_{0}^{w_{ij}}(1-x)^{k_i-2}dx < \alpha,
    \label{Eq:Edges:Alpha:ij}
\end{equation}
where $k_i$ is the number of friends avatar $i$ has. We use $\alpha=0.05$ and remove all the insignificant links, resulting a directed network. If link $i\to j$ is significant, it means that $j$ is relatively important to $i$ in $i$'s friends. Speaking differently, $i$ depends on $j$ and the directed network can be termed as a dependence network.

\begin{figure}[htb]
  \centering
  \includegraphics[width=8cm]{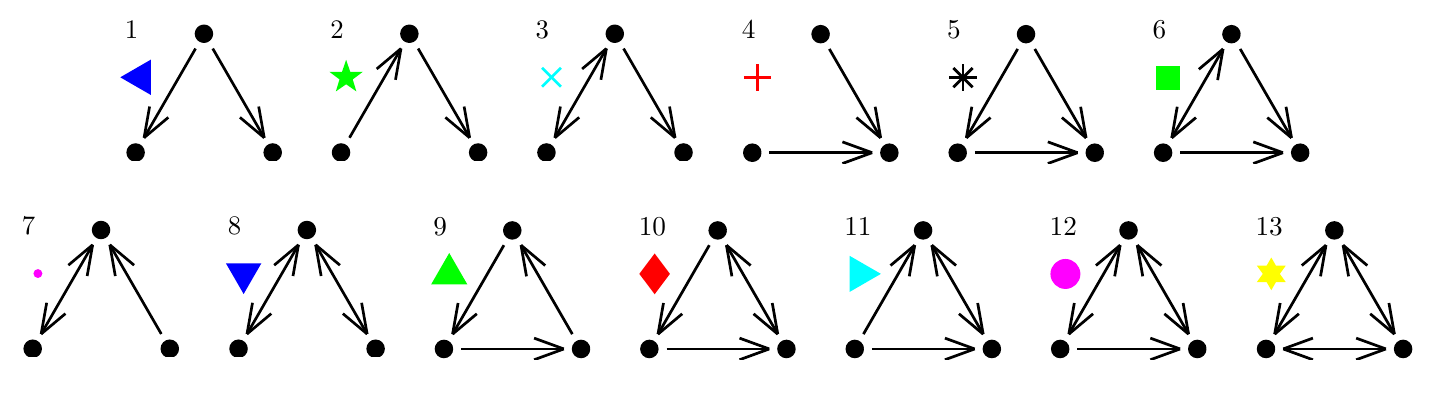}
  \caption{{\textbf{List of triadic dependence motifs.}} A unidirectional arrow from $i$ to $j$ means that avatar $i$ depends on avatar $j$, while a bidirectional arrow indicates that the two avatars are mutually dependent.}
  \label{Fig:5}
\end{figure}

\noindent{\textbf{Definition and identification of triadic motifs.}}
We consider directed triadic motifs for the dependence networks, each of which contains three connected nodes. As shown in Fig.~\ref{Fig:5}, there are 13 directed triadic motifs. These motifs represent different dependence structure among avatars at the microscopic level. For instance, motif 1 stands for the situation that one avatar depends on two other avatars, while motif 2 means that one avatar depends on another avatar which in turn depends on the third avatar. To identify motifs, we adopt a widely used method proposed in Ref.~\cite{Milo-ShenOrr-Itzkovitz-Kashtan-Chklovskii-Alon-2002-Science}.

\noindent{\textbf{Null model.}}
The specification of null models is very important in the assessment of certain statistic and the difference between the real networks and the reference randomized networks should focus on the factor under consideration \cite{Kovanen-Kaski-Kertesz-Saramaki-2013-PNAS}. Because we investigate the possible distinctions of the level sum and level difference of avatars in dependence motifs, we shuffle the levels of all avatars in dependence networks. Specifically, the topological structure of the dependence networks remains unchanged, while the levels of avatars are randomized by repeatedly exchange the levels of two arbitrarily chosen avatars.

%

\begin{thebibliography}{10}
\expandafter\ifx\csname url\endcsname\relax
  \def\url#1{\texttt{#1}}\fi
\expandafter\ifx\csname urlprefix\endcsname\relax\def\urlprefix{URL }\fi
\providecommand{\bibinfo}[2]{#2}
\providecommand{\eprint}[2][]{\url{#2}}

\bibitem{Currarini-Jackson-Pin-2009-Em}
\bibinfo{author}{Currarini, S.}, \bibinfo{author}{Jackson, M.~O.} \&
  \bibinfo{author}{Pin, P.}
\newblock \bibinfo{title}{{An economic model of friendship: homophily,
  minorities and segregation}}.
\newblock \emph{\bibinfo{journal}{Econometrica}} \textbf{\bibinfo{volume}{77}},
  \bibinfo{pages}{1003--1045} (\bibinfo{year}{2009}).

\bibitem{Currarini-Jackson-Pin-2010-PNAS}
\bibinfo{author}{Currarini, S.}, \bibinfo{author}{Jackson, M.~O.} \&
  \bibinfo{author}{Pin, P.}
\newblock \bibinfo{title}{{Identifying the roles of race-based choice and
  chance in high school friendship network formation}}.
\newblock \emph{\bibinfo{journal}{Proc. Natl. Acad. Sci. U.S.A.}}
  \textbf{\bibinfo{volume}{107}}, \bibinfo{pages}{4857--4861}
  (\bibinfo{year}{2010}).

\bibitem{Ball-Newman-2013-NS}
\bibinfo{author}{Ball, B.} \& \bibinfo{author}{Newman, M. E.~J.}
\newblock \bibinfo{title}{{Friendship networks and social status}}.
\newblock \emph{\bibinfo{journal}{Network Science}}
  \textbf{\bibinfo{volume}{1}}, \bibinfo{pages}{16--30} (\bibinfo{year}{2013}).

\bibitem{Palla-Barabasi-Vicsek-2007-Nature}
\bibinfo{author}{Palla, G.}, \bibinfo{author}{Barab{\'a}si, A.-L.} \&
  \bibinfo{author}{Vicsek, T.}
\newblock \bibinfo{title}{{Quantifying social group evolution}}.
\newblock \emph{\bibinfo{journal}{Nature}} \textbf{\bibinfo{volume}{446}},
  \bibinfo{pages}{664--667} (\bibinfo{year}{2007}).

\bibitem{Onnela-Saramaki-Hyvonen-Szabo-Lazer-Kaski-Kertesz-Barabasi-2007-PNAS}
\bibinfo{author}{Onnela, J.-P.} \emph{et~al.}
\newblock \bibinfo{title}{{Structure and tie strengths in mobile communication
  networks}}.
\newblock \emph{\bibinfo{journal}{Proc. Natl. Acad. Sci. U.S.A.}}
  \textbf{\bibinfo{volume}{104}}, \bibinfo{pages}{7332--7336}
  (\bibinfo{year}{2007}).

\bibitem{Kumpula-Onnela-Saramaki-Kaski-Kertesz-2007-PRL}
\bibinfo{author}{Kumpula, J.~M.}, \bibinfo{author}{Onnela, J.-P.},
  \bibinfo{author}{Saram{\"a}ki, J.}, \bibinfo{author}{Kaski, K.} \&
  \bibinfo{author}{Kert{\'e}sz, J.}
\newblock \bibinfo{title}{{Emergence of communities in weighted networks}}.
\newblock \emph{\bibinfo{journal}{Phys. Rev. Lett.}}
  \textbf{\bibinfo{volume}{99}}, \bibinfo{pages}{228701}
  (\bibinfo{year}{2007}).

\bibitem{Eagle-Penland-Lazer-2009-PNAS}
\bibinfo{author}{Eagle, N.}, \bibinfo{author}{Pentland, A.} \&
  \bibinfo{author}{Lazer, D.}
\newblock \bibinfo{title}{{Inferring friendship network structure by using
  mobile phone data}}.
\newblock \emph{\bibinfo{journal}{Proc. Natl. Acad. Sci. U.S.A.}}
  \textbf{\bibinfo{volume}{106}}, \bibinfo{pages}{15274--15278}
  (\bibinfo{year}{2009}).

\bibitem{Jo-Pan-Kaski-2011-PLoS1}
\bibinfo{author}{Jo, H.-H.}, \bibinfo{author}{Pan, R.~K.} \&
  \bibinfo{author}{Kaski, K.}
\newblock \bibinfo{title}{{Emergence of bursts and communities in evolving
  weighted networks}}.
\newblock \emph{\bibinfo{journal}{PLoS One}} \textbf{\bibinfo{volume}{6}},
  \bibinfo{pages}{e22687} (\bibinfo{year}{2011}).

\bibitem{Jiang-Xie-Li-Podobnik-Zhou-Stanley-2013-PNAS}
\bibinfo{author}{Jiang, Z.-Q.} \emph{et~al.}
\newblock \bibinfo{title}{{Calling patterns in human communication dynamics}}.
\newblock \emph{\bibinfo{journal}{Proc. Natl. Acad. Sci. U.S.A.}}
  \textbf{\bibinfo{volume}{110}}, \bibinfo{pages}{1600--1605}
  (\bibinfo{year}{2013}).

\bibitem{Kovanen-Kaski-Kertesz-Saramaki-2013-PNAS}
\bibinfo{author}{Kovanen, L.}, \bibinfo{author}{Kaski, K.},
  \bibinfo{author}{Kert{\'{e}}sz, J.} \& \bibinfo{author}{Saram{\"{a}}ki, J.}
\newblock \bibinfo{title}{{Temporal motifs reveal homophily, gender-specific
  patterns, and group talk in call sequences}}.
\newblock \emph{\bibinfo{journal}{Proc. Natl. Acad. Sci. U.S.A.}}
  \textbf{\bibinfo{volume}{110}}, \bibinfo{pages}{18070--18075}
  (\bibinfo{year}{2013}).

\bibitem{Bainbridge-2007-Science}
\bibinfo{author}{Bainbridge, W.~S.}
\newblock \bibinfo{title}{{The scientific research potential of virtual
  worlds}}.
\newblock \emph{\bibinfo{journal}{Science}} \textbf{\bibinfo{volume}{317}},
  \bibinfo{pages}{472--476} (\bibinfo{year}{2007}).

\bibitem{Jiang-Zhou-Tan-2009-EPL}
\bibinfo{author}{Jiang, Z.-Q.}, \bibinfo{author}{Zhou, W.-X.} \&
  \bibinfo{author}{Tan, Q.-Z.}
\newblock \bibinfo{title}{{Online-offline activities and game-playing behaviors
  of avatars in a massive multiplayer online role-playing game}}.
\newblock \emph{\bibinfo{journal}{EPL (Europhys. Lett.)}}
  \textbf{\bibinfo{volume}{88}}, \bibinfo{pages}{48007} (\bibinfo{year}{2009}).

\bibitem{Jiang-Ren-Gu-Tan-Zhou-2010-PA}
\bibinfo{author}{Jiang, Z.-Q.}, \bibinfo{author}{Ren, F.}, \bibinfo{author}{Gu,
  G.-F.}, \bibinfo{author}{Tan, Q.-Z.} \& \bibinfo{author}{Zhou, W.-X.}
\newblock \bibinfo{title}{{Statistical properties of online avatar numbers in a
  massive multiplayer online role-playing game}}.
\newblock \emph{\bibinfo{journal}{Physica A}} \textbf{\bibinfo{volume}{389}},
  \bibinfo{pages}{807--814} (\bibinfo{year}{2010}).

\bibitem{Thurner-Szell-Sinatra-2012-PLoS1}
\bibinfo{author}{Thurner, S.}, \bibinfo{author}{Szell, M.} \&
  \bibinfo{author}{Sinatra, R.}
\newblock \bibinfo{title}{{Emergence of good conduct, scaling and Zipf laws in
  human behavioral sequences in an online world}}.
\newblock \emph{\bibinfo{journal}{PLoS One}} \textbf{\bibinfo{volume}{7}},
  \bibinfo{pages}{e29796} (\bibinfo{year}{2012}).

\bibitem{Szell-Sinatra-Petri-Thurner-Latora-2012-SR}
\bibinfo{author}{Szell, M.}, \bibinfo{author}{Sinatra, R.},
  \bibinfo{author}{Petri, G.}, \bibinfo{author}{Thurner, S.} \&
  \bibinfo{author}{Latora, V.}
\newblock \bibinfo{title}{{Understanding mobility in a social petri dish}}.
\newblock \emph{\bibinfo{journal}{Sci. Rep.}} \textbf{\bibinfo{volume}{2}},
  \bibinfo{pages}{457} (\bibinfo{year}{2012}).

\bibitem{Szell-Thurner-2012-ACS}
\bibinfo{author}{Szell, M.} \& \bibinfo{author}{Thurner, S.}
\newblock \bibinfo{title}{{Social dynamics in a large-scale online game}}.
\newblock \emph{\bibinfo{journal}{Adv. Complex Sys.}}
  \textbf{\bibinfo{volume}{15}}, \bibinfo{pages}{1250064}
  (\bibinfo{year}{2012}).

\bibitem{Szell-Lambiotte-Thurner-2010-PNAS}
\bibinfo{author}{Szell, M.}, \bibinfo{author}{Lambiotte, R.} \&
  \bibinfo{author}{Thurner, S.}
\newblock \bibinfo{title}{{Multirelational organization of large-scale social
  networks in an online world}}.
\newblock \emph{\bibinfo{journal}{Proc. Natl. Acad. Sci. U.S.A.}}
  \textbf{\bibinfo{volume}{107}}, \bibinfo{pages}{13636--13641}
  (\bibinfo{year}{2010}).

\bibitem{Szell-Thurner-2010-SN}
\bibinfo{author}{Szell, M.} \& \bibinfo{author}{Thurner, S.}
\newblock \bibinfo{title}{{Measuring social dynamics in a massive multiplayer
  online game}}.
\newblock \emph{\bibinfo{journal}{Soc. Networks}}
  \textbf{\bibinfo{volume}{32}}, \bibinfo{pages}{313--329}
  (\bibinfo{year}{2010}).

\bibitem{Klimek-Thurner-2013-NJP}
\bibinfo{author}{Klimek, P.} \& \bibinfo{author}{Thurner, S.}
\newblock \bibinfo{title}{{Triadic closure dynamics drives scaling laws in
  social multiplex networks}}.
\newblock \emph{\bibinfo{journal}{New J. Phys.}} \textbf{\bibinfo{volume}{15}},
  \bibinfo{pages}{063008} (\bibinfo{year}{2013}).

\bibitem{Szell-Thurner-2013-SR}
\bibinfo{author}{Szell, M.} \& \bibinfo{author}{Thurner, S.}
\newblock \bibinfo{title}{{How women organize social networks different from
  men}}.
\newblock \emph{\bibinfo{journal}{Sci. Rep.}} \textbf{\bibinfo{volume}{3}},
  \bibinfo{pages}{1214} (\bibinfo{year}{2013}).

\bibitem{Xie-Li-Jiang-Tan-Podobnik-Zhou-Stanley-2014-PNAS}
\bibinfo{author}{Xie, W.-J.} \emph{et~al.}
\newblock \bibinfo{title}{{Division of labor, skill complementarity, and
  heterophily in socioeconomic networks}}.
\newblock \bibinfo{pages}{submitted} (\bibinfo{year}{2014}).

\bibitem{Serrano-Boguna-Vespignani-2009-PNAS}
\bibinfo{author}{Serrano, M.~{\'A}.}, \bibinfo{author}{Bogu{\~{n}}{\'a}, M.} \&
  \bibinfo{author}{Vespignani, A.}
\newblock \bibinfo{title}{{Extracting the multiscale backbone of complex
  weighted networks}}.
\newblock \emph{\bibinfo{journal}{Proc. Natl. Acad. Sci. U.S.A.}}
  \textbf{\bibinfo{volume}{106}}, \bibinfo{pages}{6483--6488}
  (\bibinfo{year}{2009}).

\bibitem{Milo-Itzkovitz-Kashtan-Levitt-ShenOrr-Ayzenshtat-Sheffer-Alon-2004-Science}
\bibinfo{author}{Milo, R.} \emph{et~al.}
\newblock \bibinfo{title}{{Superfamilies of evolved and designed networks}}.
\newblock \emph{\bibinfo{journal}{Science}} \textbf{\bibinfo{volume}{303}},
  \bibinfo{pages}{1538--1542} (\bibinfo{year}{2004}).

\bibitem{Milo-Kashtan-Itzkovitz-Newman-Alon-2004-XXX}
\bibinfo{author}{Milo, R.}, \bibinfo{author}{Kashtan, N.},
  \bibinfo{author}{Itzkovitz, S.}, \bibinfo{author}{Newman, M. E.~J.} \&
  \bibinfo{author}{Alon, U.}
\newblock \bibinfo{title}{{Uniform generation of random graphs with arbitrary
  degree sequences}} (\bibinfo{year}{2004}).
\newblock \bibinfo{note}{Http://arxiv.org/abs/cond-mat/0312028}.

\bibitem{Tran-Choi-Zhang-2013-NatComm}
\bibinfo{author}{Tran, N.~H.}, \bibinfo{author}{Choi, K.~P.} \&
  \bibinfo{author}{Zhang, L.-X.}
\newblock \bibinfo{title}{{Counting motifs in the human interactome}}.
\newblock \emph{\bibinfo{journal}{Nat. Commun.}} \textbf{\bibinfo{volume}{4}},
  \bibinfo{pages}{2241} (\bibinfo{year}{2013}).

\bibitem{Milo-ShenOrr-Itzkovitz-Kashtan-Chklovskii-Alon-2002-Science}
\bibinfo{author}{Milo, R.} \emph{et~al.}
\newblock \bibinfo{title}{{Network motifs: Simple building blocks of complex
  networks}}.
\newblock \emph{\bibinfo{journal}{Science}} \textbf{\bibinfo{volume}{298}},
  \bibinfo{pages}{824--827} (\bibinfo{year}{2002}).

\bibitem{Kovanen-Karsai-Kaski-Kertesz-Saramaki-2011-JSM}
\bibinfo{author}{Kovanen, L.}, \bibinfo{author}{Karsai, M.},
  \bibinfo{author}{Kaski, K.}, \bibinfo{author}{Kert{\'{e}}sz, J.} \&
  \bibinfo{author}{Saram{\"{a}}ki, J.}
\newblock \bibinfo{title}{{Temporal motifs in time-dependent networks}}.
\newblock \emph{\bibinfo{journal}{J. Stat. Mech.}}
  \textbf{\bibinfo{volume}{2011}}, \bibinfo{pages}{P11005}
  (\bibinfo{year}{2011}).

\bibitem{Jiang-Xie-Xiong-Zhang-Zhang-Zhou-2013-QFL}
\bibinfo{author}{Jiang, Z.-Q.} \emph{et~al.}
\newblock \bibinfo{title}{Trading networks, abnormal motifs and stock
  manipulation}.
\newblock \emph{\bibinfo{journal}{Quant. Finance Lett.}}
  \textbf{\bibinfo{volume}{1}}, \bibinfo{pages}{1--8} (\bibinfo{year}{2013}).

\end{thebibliography}

\vspace{-5mm}
\section*{Acknowledgements}
\vspace{-3mm}
\noindent{We acknowledge financial support from the National Natural Science Foundation of China (11205057 and 11375064), the Humanities and Social Sciences Fund of the Ministry of Education of China (09YJCZH040), the Ph.D. Programs Foundation of Ministry of Education of China (20120074120028), and Fundamental Research Funds for the Central Universities.}

\newpage

\end{document}